\documentclass[sigconf]{acmart}
\usepackage[noend]{algpseudocode}
\usepackage{algorithmicx,algorithm}
\usepackage{multirow}
\usepackage{graphicx}
\usepackage{enumerate}
\usepackage{enumitem}
\usepackage{balance}
\usepackage{hyperref}

\usepackage{microtype}

\AtBeginDocument{%
  \providecommand\BibTeX{{%
    \normalfont B\kern-0.5em{\scshape i\kern-0.25em b}\kern-0.8em\TeX}}}

\copyrightyear{2024}
\acmYear{2024}
\setcopyright{acmlicensed}
\acmConference[BCIMM '24] {Proceedings of the 1st International Workshop on Brain-Computer Interfaces (BCI) for Multimedia Understanding}{October 28, 2024}{Melbourne, VIC, Australia.}
\acmBooktitle{Proceedings of the 1st International Workshop on Brain-Computer Interfaces (BCI) for Multimedia Understanding (BCIMM '24), October 28, 2024, Melbourne, VIC, Australia.}
\acmISBN{979-8-4007-1189-3/24/10}
\acmDOI{10.1145/3688862.3689110}

\begin{document}
\title{Online Multi-level Contrastive Representation Distillation for Cross-Subject fNIRS Emotion Recognition}


\author{Zhili Lai}
\affiliation{%
  \institution{School of Electronic and Information Engineering, South China University of Technology}
  \city{Guangzhou}
  \country{China}}
\email{202321012469@mail.scut.edu.cn}

\author{Chunmei Qing}
\authornote{Corresponding author}
\affiliation{%
  \institution{School of Electronic and Information Engineering, South China University of Technology \& Pazhou Lab}
  \city{Guangzhou}
  \country{China}}
\email{qchm@scut.edu.cn}

\author{Junpeng Tan}
\affiliation{%
  \institution{School of Electronic and Information Engineering, South China University of Technology}
  \city{Guangzhou}
  \country{China}}
\email{tjeepscut@gmail.com}

\author{Wanxiang Luo}
\affiliation{%
  \institution{School of Electronic and Information Engineering, South China University of Technology}
  \city{Guangzhou}
  \country{China}}
\email{wanxiangluo@foxmail.com}

\author{Xiangmin Xu}
\affiliation{%
  \institution{School of Future Technology, South China University of Technology \& Pazhou Lab}
  \city{Guangzhou}
  \country{China}}
\email{xmxu@scut.edu.cn}


\begin{abstract}
\begin{sloppypar}
Utilizing functional near-infrared spectroscopy (fNIRS) signals for emotion recognition is a significant advancement in understanding human emotions. However, due to the lack of artificial intelligence data and algorithms in this field, current research faces the following challenges: 1) The portable wearable devices have higher requirements for lightweight models; 2) The objective differences of physiology and psychology among different subjects aggravate the difficulty of emotion recognition. To address these challenges, we propose a novel cross-subject fNIRS emotion recognition method, called the Online Multi-level Contrastive Representation Distillation framework (OMCRD). Specifically, OMCRD is a framework designed for mutual learning among multiple lightweight student networks. It utilizes multi-level fNIRS feature extractor for each sub-network and conducts multi-view sentimental mining using physiological signals. The proposed Inter-Subject Interaction Contrastive Representation (IS-ICR) facilitates knowledge transfer for interactions between student models, enhancing cross-subject emotion recognition performance. The optimal student network can be selected and deployed on a wearable device. Some experimental results demonstrate that OMCRD achieves state-of-the-art results in emotional perception and affective imagery tasks. 
\end{sloppypar}
\end{abstract}

\ccsdesc[500]{Computing methodologies~Artificial intelligence}
\ccsdesc[500]{Human-centered computing~HCI design and evaluation methods}

\keywords{fNIRS, Emotion Recognition, Online Knowledge Distillation, Contrastive Learning}



\maketitle

\section{Introduction}

Emotion recognition is an important research task in affective computing, which aims to represent and explain human mental states by acquiring psychological or non-psychological signals. Accurate emotion recognition can help us better sense a person's brain activity and thoughts. Given contemporary society's challenging employment and living conditions, emotion recognition technology is increasingly prevalent across diverse industries. Among them, common applications include health diagnostics, safe driving, medical services, and human-computer interaction.\cite{cowie2001emotion, waelbers2022comparing, ayata2020emotion, huang2023context,chen2024dphanet}.

\begin{sloppypar}
To this end, commonly processed data for affective computing include facial videos, body movements, speech, text scales, and physiological signals \cite{yang2018emotion, cen2024masanet, latif2021survey,  gong2023astdf}. Comparing these data sources, physiological signals come from the human brain nervous system, less influenced by subjective consciousness or deception, and offer more dependable insights into human mood shifts. Functional Near-Infrared Spectroscopy (fNIRS) stands out as a potential method for neural signal capture and visualization, to create a real-time communication link between the human brain and external devices. fNIRS signals indicate brain activity by tracking alterations in oxygenated hemoglobin (HbO) and deoxygenated hemoglobin (HbR) levels across various brain areas. Two primary features exist in the fNIRS signal: region-level features representing activation levels in different regions, and channel-level features showing the overall change in blood oxygen concentration across time.
\end{sloppypar}

While fNIRS has been crucial in measuring physiological signals, there remain numerous challenges in advancing fNIRS data research. Current deep learning-based fNIRS research primarily concentrates on distinct tasks, like Brain-Computer Interface (BCI) \cite{kwak2022fganet} and mental health diagnosis \cite{kalanadhabhatta2022extracting}, while exploration in fNIRS emotion recognition is still in the early stages of development. With the rapid development of portable wearable physiological devices, lightweight network models can maintain reasonable measurement accuracy while adapting to the resource constraints of wearable devices and application requirements. Finally, the fNIRS recordings related to emotions exhibit substantial inter-subject variabilities due to the physiological differences among individuals. This is a considerable challenge for cross-subject emotion recognition.

To address the issues mentioned above, this paper proposes a novel Online Multi-level Contrastive Representation Distillation framework for fNIRS emotion recognition, named OMCRD. We explore and experiment with fNIRS emotion recognition to improve deficiencies in this field. Specifically, the OMCRD framework employs a one-stage online multi-student network mutual Knowledge Distillation (KD) strategy with the multi-level (region-level and channel-level) fNIRS feature extractor, which exploits the mutual learning between multiple lightweight student networks during training to eliminate the dependence on complex teacher models. Furthermore, we also propose the Inter-Subject Interaction Contrastive Representation loss (IS-ICR). "Inter-Subject" means that fNIRS signals collected from different subjects facing the same stimulus are treated as the same class, facilitating the learning of their similarities. "Interaction" refers to establishing collaborative learning relationships across peer networks. Different networks can communicate with each other and optimize parameters during training. This ensures that each network learns additional contrastive representation knowledge from other peer networks. This approach facilitates the deployment of the final best-performing student model to a wearable fNIRS physiological monitoring device.

In summary, our primary contributions are as follows:

\begin{enumerate}[label=(\arabic*)] 

\item We propose a novel fNIRS emotion recognition method (OMCRD), which is an online multi-level contrastive representation distillation framework with the multiple student networks for learning multi-view fNIRS features and reducing model complexity. To our knowledge, it is the first work to apply knowledge distillation in fNIRS emotion recognition.
\item A novel Inter-Subject Interaction Contrastive Representation loss (IS-ICR) is proposed, which is a multi-level contrastive representation learning. IS-ICR effectively enables different student models to obtain region-level and channel-level knowledge across subjects and optimizes multi-view features between student models.
\item Extensive experiments are conducted on a publicly available fNIRS dataset to assess the proposed method. The results demonstrated the effectiveness and robustness of the proposed framework.
\end{enumerate}

\section{Related Work}
In this segment, to enhance comprehension of the proposed approach, we mainly introduce several related works: physiological signals emotion recognition, knowledge distillation, and contrastive learning.

\subsection{Physiological Signals Emotion Recognition}
Recently, physiological signals emotion recognition has gradually become a research hotspot in affective computing. Unlike emotion recognition of behavioral modalities \cite{yang2018emotion, piana2016adaptive, latif2021survey}, Physiological signals are challenging to disguise and cannot be intentionally or consciously controlled. Physiological signals include electroencephalogram (EEG) \cite{liu2023glfanet, gong2023astdf}, electrodermal activity (EDA) \cite{yu2020systematic}, electrocardiogram (ECG) \cite{hsu2017automatic, xiao2023spatial}, Galvanic Skin Response (GSR) \cite{susanto2020emotion}. They are frequently used in emotion recognition systems because they can mimic the real physiological changes of human emotions. Like, Xiao et al. \cite{xiao2023spatial} enhanced emotion recognition performance by acquiring spatial-temporal representations of various ECG regions and implementing a dynamic weight allocation layer to modify the influence of each ECG region. To ensure the recognition of emotional states in a long-time series, Susanto et al. \cite{susanto2020emotion} analyzed GSR signals based on a 1D convolutional neural network and a residual bidirectional gated loop unit. Meanwhile, Yu et al. \cite{yu2020systematic} investigated the performance of different deep neural networks on a subject-independent EDA-based emotion classification task. Gong et al. \cite{gong2023astdf} proposed a spatio-temporal two-stream fusion network based on an attention mechanism for EEG-based emotion recognition.

Notably, as an emerging non-invasive brain imaging technology, fNIRS has gradually gained attention in emotion recognition research due to its advantages of flexibility, ease of operation, and low cost. Si et al. \cite{si2023cross} combined the CNN branch and the statistical branch to construct a dual-branch joint network for cross-subject fNIRS emotion recognition, which is the first introduction of deep learning techniques in the field. Chen et al. \cite{chen2024temporal} presented a pioneering wearable bimodal system. The system combined fNIRS and EEG technology and used the temporal convolutional network to identify implicit emotional states in real-time. fNIRS is crucial for processing emotional brain responses and shows promise in emotion recognition. However, due to minimal AI research and challenges in feature extraction, algorithm accuracy is currently very low. Implementing a multi-student network strategy for optimizing training with multi-view features could be beneficial.

\begin{figure*}[ht]
  \centering
  \includegraphics[scale=.54]{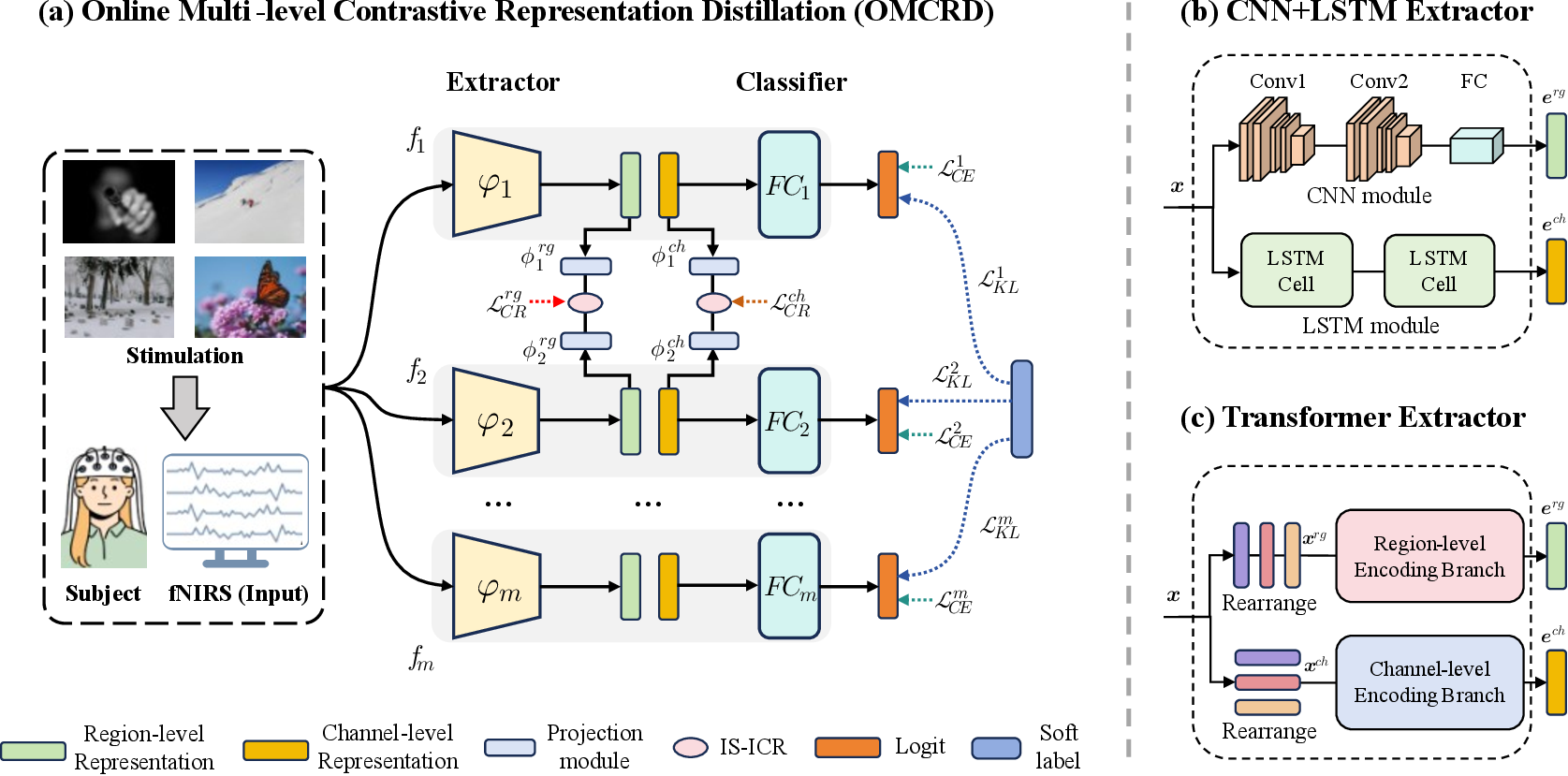}
  \caption{(a) Overview of the proposed OMCRD for fNIRS emotion recognition. The pictures of emotional stimulus sources are taken from \cite{spape2023nemo}. (b) and (c) show two different types of multi-level fNIRS feature extractors.}
  \label{fig1}
\end{figure*}

\subsection{Knowledge Distillation}

The Knowledge Distillation (KD) method was initially proposed by Hinton et al. \cite{hinton2015distilling}, focusing on distilling knowledge from a large teacher model to improve a smaller student network. KD methods facilitate the deployment of large-parameter models in practical applications. In the KD research based on physiological signals, Gu et al. \cite{gu2022frame} designed a novel frame-level teacher-student framework, achieving the best performance in subject-independent EEG emotion recognition. Wang et al. \cite{wang2022eeg} attempted to distill the knowledge from ResNet34 into a smaller model ResNet8 in EEG emotion recognition tasks to achieve performance improvement and model compression. Liu et al. \cite{liu2023emotionkd} proposed a cross-modal (GSR and EEG) knowledge distillation framework, effectively transferring heterogeneous and interactive knowledge from multimodal to unimodal GSR models, enhancing emotion recognition performance while reducing dependence on EEG signals.

However, these methods employ an offline learning strategy and require pre-training of a suitable and powerful "teacher" model. Moreover, this two-stage training process is very time-consuming, which is challenging in real-time fNIRS emotion recognition tasks. Compared to traditional KD, online KD aims to improve performance by leveraging collaborative learning across multiple student networks. Online KD employs one-stage training and does not rely on a pre-trained teacher model. Therefore, inspired by the success of online KD in the image domain \cite{zhang2018deep, zhu2018knowledge, chen2020online, wu2021peer}, we design a novel online distillation framework to ensure knowledge transfer and interaction optimization among multi-student networks.

\subsection{Contrastive Learning}

The idea of contrastive learning is to learn which data pairs are similar or different, thereby acquiring the general features of the dataset. It is a form of self-supervised learning algorithm. It has achieved outstanding performance in diverse fields, including Computer Vision (CV) \cite{chen2020simple}, Natural Language Processing (NLP) \cite{giorgi2020declutr}, and bioinformatics \cite{liu2021deep}. Khosla et al \cite{khosla2020supervised} extended contrastive loss to supervised environments, effectively using label information to extract more discriminative representations. Recently, contrastive learning has also been gradually applied to physiological signals studies. For instance, Soltanieh et al \cite{soltanieh2022analysis} investigated the effectiveness of various data augmentation techniques for contrastive self-supervised learning of ECG signals. This enables the model to enhance learning of the generalized ECG representation, thereby enhancing the accuracy of arrhythmia detection. Kalanadhabhatta et al. \cite{kalanadhabhatta2022extracting} introduced a multi-task supervised contrastive learning method to extract fNIRS, GSR, and facial video embeddings for early childhood mental disorder identification. Ensuring high-quality brain signal acquisition with portable devices involves addressing the cross-subject effects and device parameters. Shen et al. \cite{shen2022contrastive} introduced a contrastive learning-based approach for cross-subject EEG emotion recognition to enhance generalization ability. Wang et al. \cite{wang2024block} utilized contrastive learning for cross-subject cognitive workload prediction using fNIRS signals. Limited research exists on cross-subject emotion recognition in human-computer interaction based on fNIRS, with low algorithm recognition rates and deployment challenges. To address the problem of cross-subject fNIRS emotion recognition, we propose a cross-subject multi-level contrastive learning strategy. It enables the model to aggregate region-level or channel-level representations of different subjects belonging to the same class at training time, while optimizing the complementary properties between peer models.

\section{Methodology}

\subsection{Overview of the proposed framework}

\subsubsection{Notation}
\ 
\newline
\indent 
As shown in Figure \ref{fig1}, each peer/student network $f(\cdot)$ consists of a multi-level fNIRS feature extractor $\varphi(\cdot)$ and a linear classifier $FC(\cdot)$. The fNIRS signal $\boldsymbol{x}$ is mapped to the logit vector $\boldsymbol{z}$ through the network $f$, i.e. 
\begin{equation}        
    \boldsymbol{z}=FC(\varphi(\boldsymbol{x}))=f(\boldsymbol{x}).
\end{equation}
 Specifically, the input $\boldsymbol{x}$ is processed by the extractor $\varphi$ to obtain the region-level embedding $\boldsymbol{e}^{rg}$ and the channel-level embedding $\boldsymbol{e}^{ch}$, which are concatenated and then fed into $FC$. For the intermediate feature embeddings $\boldsymbol{e}^{rg}$ and $\boldsymbol{e}^{ch}$, we introduce two projection modules $\phi^{rg}(\cdot)$ and $\phi^{ch}(\cdot)$ respectively. The process is as follows:
 \begin{equation}        
    \boldsymbol{v}^{rg}=\phi^{rg}(\boldsymbol{e}^{rg})
\end{equation}
 \begin{equation} 
    \boldsymbol{v}^{ch}=\phi^{ch}(\boldsymbol{e}^{ch})
\end{equation}
 where $\phi$ is composed of a linear layer and the $l_{2}$-normalization to linearly transform these embeddings into contrastive embeddings $\boldsymbol{v}^{rg}$, $\boldsymbol{v}^{ch}\in\mathbb{R}^d$, and $d$ is the embedding size. The embeddings $\boldsymbol{v}^{rg}$ and $\boldsymbol{v}^{ch}$ are used for the calculation of the proposed IS-ICR.

\subsubsection{Training and deployment}
\ 
\newline
\indent 
During training, $M$ $(M\geqslant2)$ peer networks are optimized together. Notably, to learn diverse representations, all identical networks $\{f_{m}\}_{m=1}^{M}$ are assigned different weights initially, a crucial factor for mutual learning success. In the evaluation phase, projection modules $\phi$ are omitted, allowing the assessment of each network independently. The optimal performing network can be selected for final deployment. As the structure of the preserved network is identical to the others, there are no extra computational expenses during testing.

\subsection{Multi-level fNIRS Feature Extractor}
The fNIRS signal has two important features: region-level features and channel-level features. Region-level features represent the spatial correlation between different fNIRS signal channels. Over a while, changes in blood oxygen concentration from different channels often exhibit different dependencies, such as the higher correlation between channels from similar functional brain regions or neighboring spatial locations. Channel-level features represent the temporal continuity of blood oxygen concentration changes within multiple fNIRS signal channels. We designed two feature extractors to capture these two types of features. One is a hybrid feature extractor based on CNN+LSTM, and the other is a feature extractor based on Transformers.

\subsubsection{CNN+LSTM extractor}
\ 
\newline
\indent 
For the hybrid feature extractor with CNN and LSTM, we design a CNN module $\varphi^{CNN}(\cdot)$ to extract region-level features $\boldsymbol{e}^{rg}$. The $\varphi^{CNN}$ mainly consists of two 1D convolutional layers (with kernel sizes of 50 and 10, strides of 10 and 2, and output channels of 32 and 16, respectively) and a fully connected layer. Simultaneously, we also design an LSTM module $\varphi^{LSTM}(\cdot)$ containing two LSTM layers (with 64 hidden units) to extract channel-level features $\boldsymbol{e}^{ch}$. The formula is defined as follows:
\begin{equation}
    \boldsymbol{e}^{rg}=\varphi^{CNN}(\boldsymbol{x})
\end{equation}
\begin{equation}
    \boldsymbol{e}^{ch}=\varphi^{LSTM}(\boldsymbol{x})
\end{equation}
The resulting representation vectors $\boldsymbol{e}^{rg}$ and $\boldsymbol{e}^{ch}$ serve as inputs to the linear classifier. It is worth noting that to accommodate the CNN+LSTM feature extractor, the input is $\boldsymbol{x}\in\mathbb{R}^{2n\times T}$, where $n$ is the number of HbO (HbR) channels and $T$ is the signal length.

\subsubsection{Transformer extractor}
\ 
\newline
\indent 
For the Transformer-based feature extractor, with $\boldsymbol{x}\in\mathbb{R}^{2\times n\times T}$ as input. Inspired by iTransformer \cite{liu2023itransformer}, each fNIRS channel's complete sequence serves as a token for region-level modeling. Moreover, building on the idea in Informer \cite{zhou2021informer}, the information from multiple fNIRS channels at the same timestamp acts as a token for capturing channel-level feature information. 

\begin{figure}[ht]
  \centering
  \includegraphics[width=0.85\linewidth]{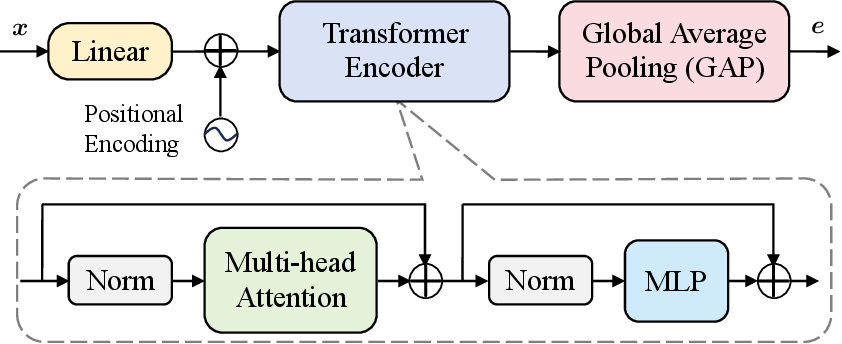}
  \caption{The architecture of the encoding branch unit in the Transformer extractor.}
  \label{fig2}
\end{figure}

Specifically, the input $\boldsymbol{x}$ is first reshaped into $\boldsymbol{x}^{rg}\in\mathbb{R}^{n\times (2\cdot T)}$ and $\boldsymbol{x}^{ch}\in\mathbb{R}^{T\times (2\cdot n)}$. Then $\boldsymbol{x}^{rg}$ and $\boldsymbol{x}^{ch}$ are fed into the region-level encoding branch $\varphi^{REB}(\cdot)$ and the channel-level encoding branch $\varphi^{CEB}(\cdot)$, respectively. The process is as follows:
\begin{equation}
    \boldsymbol{e}^{rg}=\varphi^{REB}(\boldsymbol{x}^{rg})
\end{equation}
\begin{equation}
    \boldsymbol{e}^{ch}=\varphi^{CEB}(\boldsymbol{x}^{ch})
\end{equation}
Both encoding branches consist of a linear projection layer, positional encoding, the Transformer encoder \cite{vaswani2017attention}, and a Global Average Pooling (GAP) layer \cite{lin2013network} (see figure \ref{fig2} for details).

\subsection{Learning Objectives}
Given a sample set $\mathcal{D}=\{(\boldsymbol{x}_i,y_i)\}_{i=1}^N$ containing $N$ instances (sample from different subjects) from $C$ classes, where $y_i\in\{1,2,...,C\}$. $\mathcal{D}$ as the input to $M$ networks.

\subsubsection{Learning from labels}
\ 
\newline
\indent 
Each network is optimized by Cross-Entropy loss (CE) between probability distributions and hard labels. The probability of class $c$ for sample $\boldsymbol{x}_i$ given by the $m$-th network is calculated as:
\begin{equation}
    p_m^c(\boldsymbol{x}_i)={exp(z_m^{i,c})}/{[\sum_{c=1}^Cexp(z_m^{i,c})]}
\end{equation}
where the logit $\boldsymbol{z}_m$ from $f_m$ is the input of the "softmax" layer. Therefore, the CE loss of the $m$-th network is computed as:
\begin{equation}
        \mathcal{L}_{CE}^m=-\frac{1}{N} \sum_{i=1}^N\sum_{c=1}^CI_{y_i=c}\log\left(p_m^c(\boldsymbol{x}_i)\right)
\end{equation}
where $I_{y_{i}= c}\in\{0,1\}$ represents the indicator function, which takes 1 when $y_{i}= c$ holds, and 0 otherwise. Overall, CE loss of $M$ networks is:
\begin{equation}
    \mathcal{L}_{CE}=\sum_{m=1}^M\mathcal{L}_{CE}^m
\label{CE-loss}
\end{equation}

\subsubsection{Distillation from soft labels}
\ 
\newline
\indent 
We simply construct an "online teacher" by softening the true label with temperature $T$. The probability distribution of the soft label contains higher quality logit knowledge. It is as follows:
\begin{equation}
    \tilde{p}^c(y_i)={exp(y^{i,c}/T)}/{[\sum_{c=1}^Cexp(y^{i,c}/T)]}
\end{equation}
where $y^{i,c}\in\{0,1\}$ is the $c$-th element in the one-hot encoding $\boldsymbol{y}_i$ of label $y_i$. The soft probability of class $c$ corresponding to sample $\boldsymbol{x}_i$ from network $f_m$ is:
\begin{equation}
    \tilde{p}_m^c(\boldsymbol{x}_i)={exp(z_m^{i,c}/T)}/{[\sum_{c=1}^Cexp(z_m^{i,c}/T)]}
\end{equation}
To measure the alignment between the predictions of the network $f_m$ and the soft labels, while also improving generalization, we adopt the Kullback Leibler (KL) Divergence. The KL loss of $f_m$ is calculated as follows:
\begin{equation}
\mathcal{L}_{KL}^m=D_{KL}\left(\tilde{p}^c(y_i)\|\tilde{p}_m^c(\boldsymbol{x}_i)\right)=\frac{1}{N}\sum_{i=1}^N\sum_{c=1}^C\tilde{p}^c(y_i)\log\frac{\tilde{p}^c(y_i)}{\tilde{p}_m^c(\boldsymbol{x}_i)}
\end{equation}
Thus, the total KL loss of $M$ networks is calculated as follows:
\begin{equation}
    \mathcal{L}_{KL}=\sum_{m=1}^M\mathcal{L}_{KL}^m
\label{KL-loss}
\end{equation}

\subsubsection{Inter-subject interaction contrastive representation loss}
\begin{figure}[ht]
  \centering
  \includegraphics[width=0.93\linewidth]{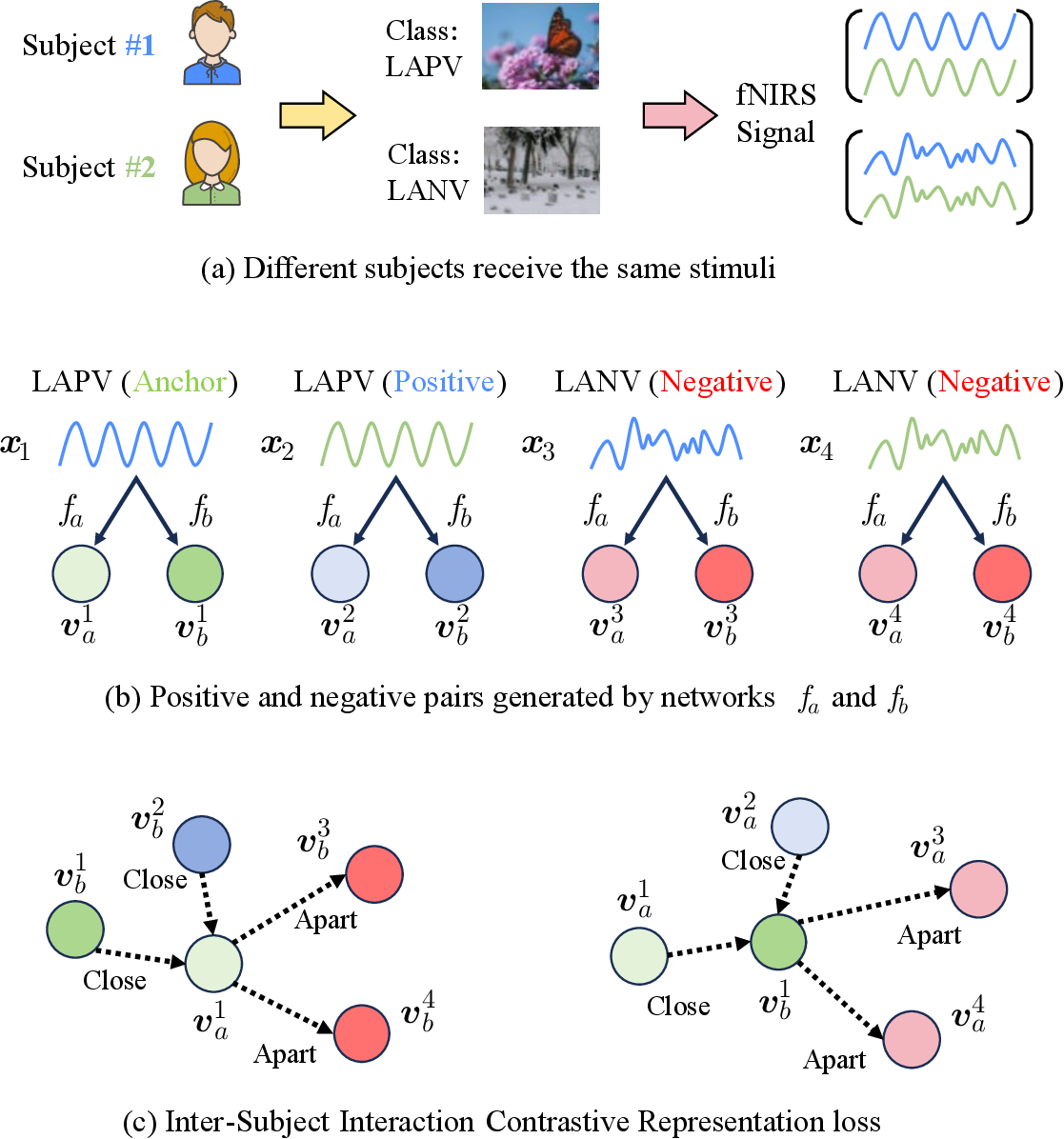}
  \caption{Overview of the proposed IS-ICR. $f_a$ and $f_b$ represent two different sub-networks in the framework. $\boldsymbol{v}_m^i$ is the feature embedding vector of the input instance $\boldsymbol{x}_i$ obtained through the network $f_m$.}
  \label{fig3}
\end{figure}

\ 
\newline
\indent 
During network training, contrastive learning techniques can push positive pairs from the same class as the anchor closer together in the embedding space, while pushing negative pairs from different classes further apart. To this end, we propose a novel Inter-Subject Interaction Contrastive Representation Loss (IS-ICR). This loss is designed to enhance the representation similarity between samples of the same class from different subjects (as shown in Figure \ref{fig3}) while capturing the general distribution of the same class of emotions. Notably, IS-ICR also makes full use of the information interaction between various peer networks to learn better representations through cross-network interaction.

Given two networks $f_a$ and $f_b$ for illustration, where $a,b\in\{1,2,\cdots,M\},a\neq b$. The embeddings generated from $\mathcal{D}$ are $\{\boldsymbol{v}_a^i\}_{i=1}^N$ and $\{\boldsymbol{v}_b^j\}_{j=1}^N$, respectively. From the perspective of $f_a$ to $f_b$ (as shown in Figure \ref{fig3}), given the anchor embedding $v_a^{1}$ of instance $x^{1}$ from $f_a$, and the embeddings in the contrastive embeddings $\{\boldsymbol{v}_b^j\}_{j=1}^N$ that is of the same class as $v_a^{1}$ are considered as positive embeddings, otherwise as negative embeddings. Since the embeddings are preprocessed through $l_{2}$-normalization, the dot product measures the similarity distribution between the anchor and contrastive embeddings. The formula is defined as follows:
\begin{equation}
    \mathcal{L}_{CR}^{a\to b}=-\sum_{i=1}^N\frac{1}{N_{y_i}}\sum_{j=1}^NI_{y_i=y_j}\log\left[\frac{\exp(v_a^i \cdot v_b^j/\tau)}{\exp(v_a^i \cdot v_b^j/\tau)+Z_i}\right]
\label{CR-loss}
\end{equation}
\begin{equation}
    Z_i=\sum_{k=1}^NI_{y_i\neq y_k}\exp(v_a^i \cdot v_b^k/\tau)
\label{Z_i-loss}
\end{equation}
where $y_j$, $y_k$ denote the labels of anchor sample $x_i$ from $f_a$ and contrastive samples $x_j$ and $x_k$ from $f_b$, respectively. $N_{y_i}$ represents the number of samples whose label is $y_i$ in $\mathcal{D}$. $\tau$ is the temperature coefficient. Intuitively, each network can benefit from Eq.(\ref{CR-loss}) by gaining additional comparative knowledge from other networks. Generic inter-subject fNIRS representations that capture specific emotions using network clusters can help generalize to unseen emotional stimuli or unseen subjects to predict emotional categories. When scaling to $M$ networks, the overall loss function can be listed as follows:
\begin{equation}
    \mathcal{L}_{CR}=\sum_{1\leq a<b\leq M}^M\left(\mathcal{L}_{CR}^{a\to b}+\mathcal{L}_{CR}^{b\to a}\right)
\end{equation}
Considering region-level and channel-level representations, two losses $\mathcal{L}_{CR}^{rg}$ and $\mathcal{L}_{CR}^{ch}$ are obtained.

\begin{algorithm}[t]
\caption{\textit{The Proposed OMCRD Algorithm}} 
\label{algo1}
\hspace*{0.02in}\textbf{Input:} Training data $\mathcal{D}=\{(\boldsymbol{x}_i,y_i)\}_{i=1}^N$ from different subjects.\\
\hspace*{-0.08in}\textbf{Output:} Trained models $\{f_{m}\}_{m=1}^{M}$ with parameters $\{\theta_{m}\}_{m=1}^{M}$.
\begin{algorithmic}[1]
\State \text{/$\ast$\textit{Training}$\ast$/}
\State \textbf{Initialisation}: Randomly initialise parameters
\For{1 $\to Epoch_{max}$} 
    \State Obtain the embeddings and logits of each peer network;
    \State Soften labels and probability distributions;
    \State Compute the classification loss $\mathcal{L}_{CE}$ (Eq.(\ref{CE-loss})); 
    \State Compute the distillation loss $\mathcal{L}_{KL}$ (Eq.(\ref{KL-loss}));
    \State Compute two contrastive losses $\mathcal{L}_{CR}^{rg}$ and $\mathcal{L}_{CR}^{ch}$(Eq.(\ref{CR-loss}));
    \State Compute the final loss function (Eq.(\ref{total-loss}));
    \State Update the parameters $\{\theta_{m}\}_{m=1}^{M}$ by the AdamW optimizer.
\EndFor
\State \textbf{end for}
\State \text{/$\ast$\textit{Testing}$\ast$/}
\State Select the best performing model in $\{f_{m}\}_{m=1}^{M}$ for deployment.
\end{algorithmic}
\end{algorithm}

\subsubsection{Overall loss}
\ 
\newline
\indent 
The overall loss function $\mathcal{L}$ is a weighted sum of the loss terms mentioned above, which is illustrated as follows:
\begin{equation}
    \mathcal{L} = \mathcal{L}_{CE} + T^2\mathcal{L}_{KL} + \alpha\mathcal{L}_{CR}^{rg} + \beta\mathcal{L}_{CR}^{ch}
\label{total-loss}
\end{equation}
where $T^2$ is used to balance the contribution of soft labels to the total loss, and $\alpha$ and $\beta$ are the weighting coefficients for the region-level and channel-level contrastive representation losses, respectively.

The entire algorithmic process is shown in Algorithm \ref{algo1}. It is worth noting that OMCRD follows a one-stage training approach and does not require the pre-trained teacher model.

\begin{figure}[t]
  \centering
  \includegraphics[width=0.75\linewidth, trim=0 80 0 0, clip]{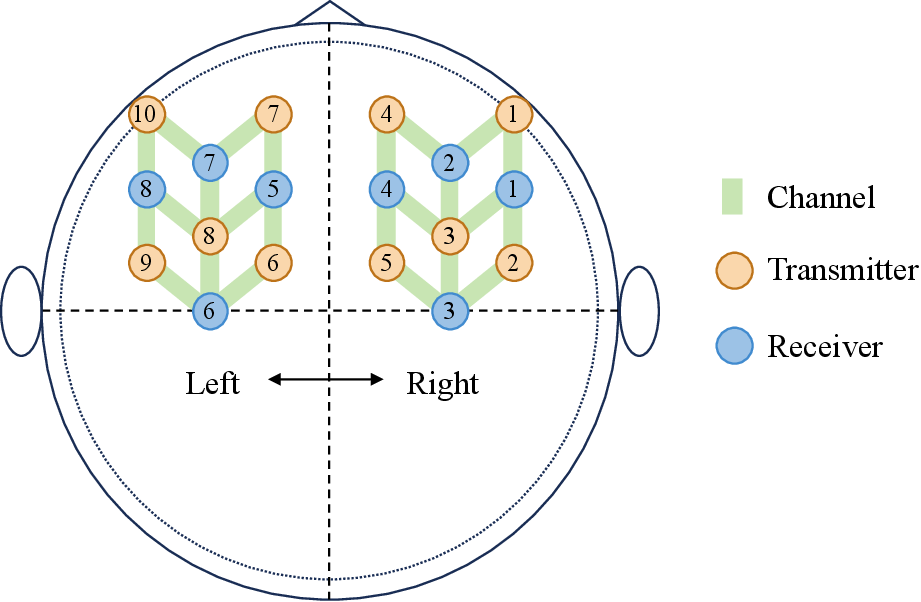}
  \caption{The channel location of fNIRS. Orange circles are 10 transmitters, blue circles are 8 receivers, and green lines are 24 fNIRS channels.}
  \label{fig4}
\end{figure}

\section{Experiments}

\subsection{Implementation Details}
\subsubsection{Datasets}
\ 
\newline
\indent 
\begin{sloppypar}
We conducted fNIRS emotion analysis on the NEMO dataset \cite{spape2023nemo}. The dataset records the brain activity of 31 subjects during two tasks: Emotional perception (Empe) and Affective imagery (Afim). For Empe, subjects viewed images from the International Affective Picture System (IAPS) database \cite{lang2007international} passively. Afim participants evaluated emotional scenarios based on textual descriptions for subjective valence and arousal levels. As shown in Figure \ref{fig4}, The fNIRS recordings were collected using a 24-channel device at 50 Hz sampling rate and grouped into four categories based on valence and arousal. High-Arousal Positive-Valence (HAPV), and High-Arousal Negative-Valence (HANV), Low-Arousal Positive-Valence (LAPV), Low-Arousal Negative-Valence (LANV). The dataset contains 1203 records for the Empe task and 720 records for the Afim task, as detailed in Table \ref{tab:dataset}.
\end{sloppypar}

        

\begin{table}
  \caption{Statistics of datasets.}
  \label{tab:dataset}
    \begin{tabular}{ccccc}
    \toprule
        \multirow{2.5}{*}{\textbf{Class}} & \makebox[0.06\textwidth][c]{\multirow{2.5}{*}{\textbf{Channels}}} & \multirow{2.5}{*}{\textbf{Length}} & \multicolumn{2}{c}{\textbf{Records}} \\
        \cmidrule(lr){4-5}
        ~ & ~ & ~ & \textbf{Empe} & \textbf{Afim} \\
        \midrule
        HAPV & 24 & 600 & 301 & 180\\
        HANV & 24 & 600 & 300 & 180\\
        LAPV & 24 & 600 & 301 & 180\\
        LANV & 24 & 600 & 301 & 180\\
        
    \bottomrule 
    \end{tabular}
\end{table}

\subsubsection{Training details}
\ 
\newline
\indent 

The dataset of around 31 participants (subjects numbered 26 and 41 completed only half of the experiment) is divided into training and testing subsets following an 8:2 proportion, iterated 5 times. To simulate real-world application scenarios, the samples used in the testing phase should come from subjects not seen during the training phase. We train the models using the AdamW \cite{loshchilov2018decoupled} optimizer with decay parameters $\beta_{1}$=0.9 and $\beta_{2}$=0.999, and chose cosine learning rate decay \cite{loshchilov2016sgdr} as the learning rate scheduler. To improve the model's generalization ability, we adopt a label smoothing \cite{muller2019does} strategy. For the architecture that includes only 2 peer networks, we trained for 60 epochs and defined the weight decay as 2. For the architecture that provides for more than 2 peer networks, epochs is 90 and the weight decay is 3. $\tau$, $\alpha$, and $\beta$ are set as 0.1, 0.2, and 0.2. For the Empe and Afim tasks, $T$ is set to 2 and 5, respectively. The learning rates for the CNN+LSTM and Transformer feature extractors are set to 0.00005 and 0.0002, respectively. The batch size for all experiments is 64. We use a specific sampler to ensure that each batch contains 16 samples from each class (4 classes), which is key to the effectiveness of IS-ICR. Our model is implemented using Python 3.10 and PyTorch 2.1, and it is trained on an NVIDIA RTX 3080Ti (16GB).

\begin{table}
  \caption{Comparison of online knowledge distillation methods based on CNN+LSTM extractor.}
  \label{tab:cnn+lstm}
  \resizebox{\linewidth}{!}{
    \begin{tabular}{llcccccc}
    \toprule
        \multirow{2.5}{*}{\textbf{Datasets}} & \multirow{2.5}{*}{\textbf{Methods}} & \multicolumn{5}{c}{\textbf{Testing Conditions}} & ~ \\
        \cmidrule(lr){3-7}
        ~ & ~ & Fold 1 & Fold 2 & Fold 3 & Fold 4 & Fold 5 & \textbf{Avg.}\\
        \midrule
        \multirow{5.5}{*}{\textbf{Empe}} & Baseline & 30.83 & 32.31 & 28.75 & 31.67 & 32.08 & 31.13 \\
        \rule{0pt}{9pt} & DML \cite{zhang2018deep} & \underline{32.92} & 34.16 & 31.25 & 31.25 & 32.50 & 32.42 \\
        \rule{0pt}{9pt}& KDCL \cite{guo2020online} & 30.83 & \underline{34.57} & 31.25 & 31.67 & \underline{34.00} & \underline{32.46} \\
        \rule{0pt}{9pt}& MCL \cite{yang2022mutual} & 31.25 & 32.10 & \underline{32.08} & \underline{32.80} & 32.50 & 32.15 \\
        \rule{0pt}{9pt}& \textbf{Ours} & \textbf{33.33} & \textbf{34.98} & \textbf{33.75} & \textbf{34.17} & \textbf{34.17} & \textbf{34.08} \\
        \midrule
        
        \multirow{5.5}{*}{\textbf{Afim}} & Baseline & 31.25 & 32.64 & 31.94 & 29.86 & 32.60 & 31.66 \\
        \rule{0pt}{9pt} & DML \cite{zhang2018deep} & 34.03 & 32.94 & 32.64 & \underline{34.03} & 31.25 & 32.98 \\
        \rule{0pt}{9pt}& KDCL \cite{guo2020online} & \underline{35.03} & \textbf{34.03} & \underline{34.72} & 31.94 & 31.94 & \underline{33.53} \\
        \rule{0pt}{9pt}& MCL \cite{yang2022mutual} & 31.94 & \underline{33.75} & \underline{34.72} & 32.64 & \underline{32.64 }& 33.14 \\
        \rule{0pt}{9pt}& \textbf{Ours} & \textbf{35.42} & \textbf{34.03} & \textbf{35.42} & \textbf{36.11} & \textbf{34.03} & \textbf{35.00} \\
    \bottomrule 
    \end{tabular}
    }
\end{table}

\begin{table}
  \caption{Comparison of online knowledge distillation methods based on Transformer extractor.}
  \label{tab:tf}
  \resizebox{\linewidth}{!}{
    \begin{tabular}{llcccccc}
    \toprule
        \multirow{2.5}{*}{\textbf{Datasets}} & \multirow{2.5}{*}{\textbf{Methods}} & \multicolumn{5}{c}{\textbf{Testing Conditions}} & ~ \\
        \cmidrule(lr){3-7}
        ~ & ~ & Fold 1 & Fold 2 & Fold 3 & Fold 4 & Fold 5 & \textbf{Avg.}\\
        \midrule
        \multirow{5.5}{*}{\textbf{Empe}} & Baseline & 31.67 & 31.28 & 32.52 & 31.67 & 33.75 & 32.18 \\
        \rule{0pt}{9pt} & DML \cite{zhang2018deep} & 33.75 & 31.69 & \underline{32.92} & \underline{32.92} & 32.50 & 32.76 \\
        \rule{0pt}{9pt}& KDCL \cite{guo2020online} & \underline{34.00} & 32.10 & 32.50 & 32.50 & \underline{34.17} & 33.05 \\
        \rule{0pt}{9pt}& MCL \cite{yang2022mutual} & \textbf{34.17} & \underline{32.51} & 32.50 & \underline{32.92} & 33.75 & \underline{33.17} \\
        \rule{0pt}{9pt}& \textbf{Ours} & \textbf{34.17} & \textbf{34.16} & \textbf{33.75} & \textbf{34.11} & \textbf{35.00} & \textbf{34.24} \\
        \midrule
        
        \multirow{5.5}{*}{\textbf{Afim}} & Baseline & 30.56 & 29.56 & 31.94 & 26.39 & 29.86 & 29.66 \\
        \rule{0pt}{9pt} & DML \cite{zhang2018deep} & 32.64 & 35.10 & \underline{36.00} & 27.08 & 29.86 & 32.14 \\
        \rule{0pt}{9pt}& KDCL \cite{guo2020online} & 31.94 & 34.03 & 35.81 & 29.17 & \underline{30.18} & 32.23 \\
        \rule{0pt}{9pt}& MCL \cite{yang2022mutual} & \underline{34.72} & \underline{35.42} & 34.72 & \underline{30.11} & 29.17 & \underline{32.83} \\
        \rule{0pt}{9pt}& \textbf{Ours} & \textbf{35.42} & \textbf{36.00} & \textbf{36.11} & \textbf{30.56} & \textbf{31.94} & \textbf{34.01} \\
    \bottomrule 
    \end{tabular}
    }
\end{table}

\subsection{Quantitative Results and Analysis}
To demonstrate the effectiveness of the proposed OMCRD, we compare it with several classic online distillation frameworks, including DML \cite{zhang2018deep}, KDCL \cite{guo2020online}, and MCL \cite{yang2022mutual}. Unless stated otherwise, we adopt the 3 peer networks ($m$ = 3) design. All frameworks are experimented with several times based on CNN+LSTM and Transformer feature extractors. The experimental results are all evaluated in terms of accuracy, with bold results indicating the best and underlined results representing the second-best results. As shown in Tables \ref{tab:cnn+lstm} and \ref{tab:tf}, it is evident that the proposed OMCRD achieves the best results in emotion recognition across different tasks.

\begin{table}
  \caption{Computational performance and parameter cost results. \# Compress denotes the compression ratio.}
  \label{tab:compression}
  \begin{tabular}{lcccc}
    \toprule
        \multirow{2.5}{*}{\textbf{Metric}} & \multicolumn{2}{c}{CNN+LSTM} & \multicolumn{2}{c}{Transformer} \\
        \cmidrule(lr){2-3} \cmidrule(lr){4-5}
        ~ & Training & Inference & Training & Inference \\
    \midrule
        \rule{0pt}{8pt} Param. (K) & 515.29 & 165.49 & 306.73 & 95.97\\
        \rule{0pt}{9pt} MACs (M) & 13.3 & 4.43 & 36.21 & 12.07\\
        \rule{0pt}{9pt} FLOPs (M) & 250.11 & 83.37 & 73.25 & 24.42\\
    \hline
        \rule{0pt}{9pt} \# Compress & \multicolumn{2}{c}{67.88\%} & \multicolumn{2}{c}{68.71\%}\\
    \bottomrule
    \end{tabular}
\end{table}

\begin{sloppypar}
The experimental results for both feature extractors yield three main conclusions. Due to mutual learning among networks, all frameworks show improvement compared to the baseline (peer networks are independent of each other). For example, the CNN+LSTM-based DML, despite its relatively poor performance, still sees an average improvement of 1.29\% and 1.32\% on the Empe and Afim. Secondly, various online knowledge distillation frameworks exhibit different adaptability levels with different extractors. With the Transformer extractor, MCL achieves suboptimal average accuracy rates of 33.17\% and 32.83\% on the Empe and Afim tasks, respectively, but fails to achieve the desired performance on the CNN+LSTM extractor. 
\end{sloppypar}

Thirdly, OMCRD excels across various feature extractors and datasets, showcasing the effectiveness and robustness of the proposed framework. Specifically, OMCRD proposes a contrastive representation learning method for cross-subject generalization, which employs a multi-peer model to capture similar features across subjects for the same emotion. Alignment of similar emotion distributions is encouraged through cross-network learning. In contrast, MCL and KDCL focus only on the knowledge of logit distribution, and these methods ignore the information of intermediate features. Suffering from incomplete information transfer, their classification performance is relatively low. For DML, each peer network learns from each other by aligning the output with other peer networks. For KDCL, the original inputs are first augmented with different random seeds separately to enhance the network's invariance to input perturbations. The predictions of the peer models are also effectively integrated to generate a soft target to ensure that multiple sub-networks benefit from collaborative learning. In addition, MCL proposes Interactive Contrastive Learning (ICL) based on Vanilla Contrastive Learning (VCL). ICL maximizes the lower bound on the mutual information between two networks by consolidating inter-network representation information. Meanwhile, distributions can be considered as soft labels to guide other distributions. However, MCL ignores the fNIRS region-level and channel-level representation information between subjects, which limits its performance. In summary, the proposed OMCRD improves the performance over other online distillation frameworks by approximately 1.1\%-1.5\%. Therefore, OMCRD demonstrates high performance and robustness in sentiment recognition tasks.

Furthermore, we assess the OMCRD based on parameters, computational speed, and compression ratio. Results in Table \ref{tab:compression} show metrics for training and inference phases. Experiments reveal faster inference computation with compression ratios of 67.88\% and 68.71\%. These findings highlight the framework's ability to compress models through online knowledge distillation, enhancing performance on wearable devices in resource-constrained settings.

\begin{table}
  \caption{Results of ablation study on loss terms.}
  \label{tab:loss ablation}
  \begin{tabular}{lcccc}
    \toprule
        \multirow{2.5}{*}{\textbf{Variants}} & \multicolumn{2}{c}{CNN+LSTM} & \multicolumn{2}{c}{Transformer} \\
        \cmidrule(lr){2-3} \cmidrule(lr){4-5}
        ~ & Empe & Afim & Empe & Afim \\
    \midrule
        \rule{0pt}{8pt} Baseline & 31.13 & 31.66 & 32.18 & 29.66\\
        \rule{0pt}{9pt} w/o \small$\mathcal{L}_{KL}$, \small$\mathcal{L}_{CR}^{ch}$ & 32.83 & 33.20 & 32.89 & 32.55\\
        \rule{0pt}{9pt} w/o \small$\mathcal{L}_{KL}$, \small$\mathcal{L}_{CR}^{rg}$ & 32.92 & 32.98 & 33.26 & 32.64\\
        \rule{0pt}{9pt} w/o \small$\mathcal{L}_{KL}$ & 33.08 & 34.28 & 33.75 & 32.98\\
        \rule{0pt}{9pt} \textbf{Completed} & \textbf{34.08} & \textbf{35.00} & \textbf{34.24} & \textbf{34.01}\\
    \bottomrule
    \end{tabular}
\end{table}

\subsection{Ablation Studies}
\subsubsection{Impact of loss items.}
\ 
\newline
\indent 
We conduct additional ablation experiments for two tasks on the Nemo dataset to understand better the contribution of different loss terms in our proposed framework. Specifically, several different variants are designed in the same experimental setup:
\begin{itemize}
\setlength{\parskip}{2pt}

\item {w/o \small$\mathcal{L}_{KL}$, \small$\mathcal{L}_{CR}^{ch}$}:  Using only region-level representations.
\item {w/o \small$\mathcal{L}_{KL}$, \small$\mathcal{L}_{CR}^{rg}$}: Using only channel-level representations.
\item {w/o \small$\mathcal{L}_{KL}$}: Distillation without logits knowledge.
\end{itemize}

As shown in Table \ref{tab:loss ablation}, each loss term in the proposed OMCRD is indispensable. These loss terms convey valuable logits or feature knowledge. The peer models learn high-quality soft label knowledge through $\mathcal{L}_{KL}$. $\mathcal{L}_{CR}^{rg}$ and $\mathcal{L}_{CR}^{ch}$ effectively transfer multi-level representations information of fNIRS signals between different models, which helps information interaction between peers.

\begin{figure}[ht]
  \centering
  \includegraphics[width=0.92\linewidth]{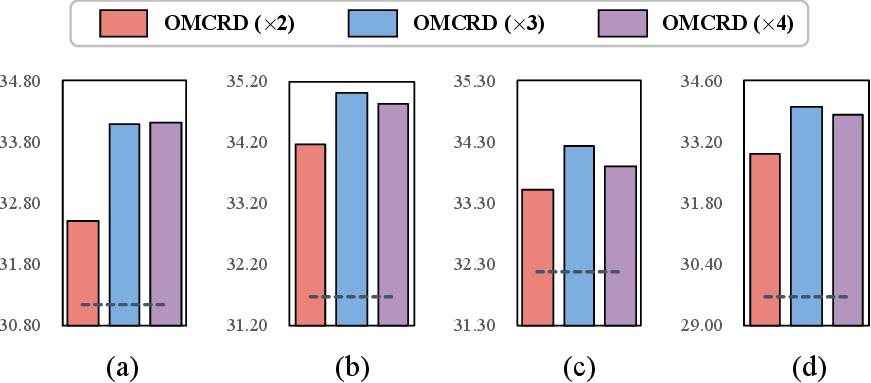}
  \caption{Ablation study results with varying numbers of peers. (a) and (b) show the results of OMCRD based on the CNN+LSTM extractor on the Empe and Afim tasks. (c) and (d) show the results based on the Transformer extractor. Dashed lines indicate baseline results}
  \label{fig5}
\end{figure}

\subsubsection{Number of peers.}
\ 
\newline
\indent
Figure \ref{fig5} illustrates that combining several peer networks can markedly surpass the basic model's performance. This outcome confirms our hypothesis that sharing contrastive knowledge among peer networks enhances overall generalization ability. Nonetheless, with an increasing number of peers in the queue, the improvement diminishes gradually, reaching a point of stabilization or even demonstrating negative growth. The main experiment has produced impressive results using three student networks.

\subsection{Visualization}
\begin{figure}[ht]
  \centering
  \includegraphics[width=0.95\linewidth]{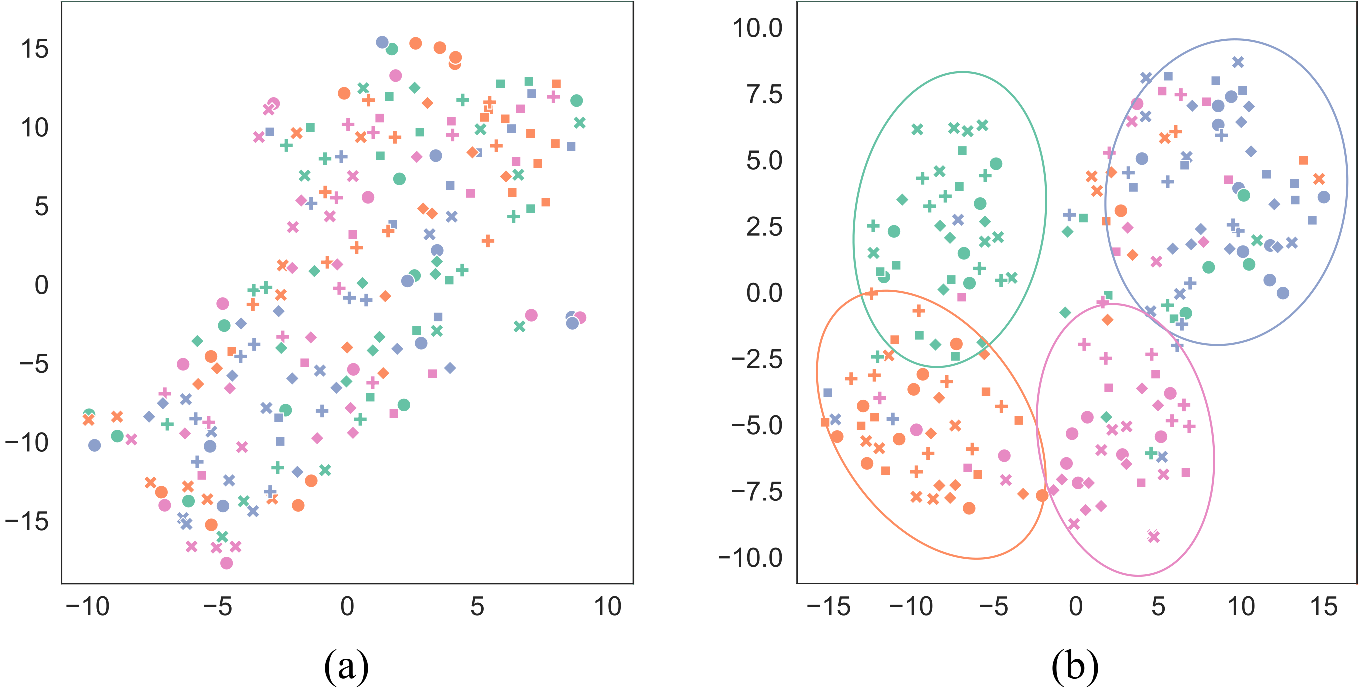}
  \caption{(a) t-SNE results for the original data. (b) t-SNE results of OMCRD (using fold 1 data). Different colors are classes: Green for LANV, Orange for HANV, Blue for LAPV, and Pink for HAPV. Different markers are various subjects.}
  \label{fig6}
\end{figure}

To further illustrate the effectiveness of the proposed OMCRD, we visualize the raw and trained features of five subjects through t-SNE (Figure \ref{fig6}). We represented different emotion categories with four colors and used five shapes to distinguish samples with different subjects. Figure \ref{fig6} (a) shows the scattered distribution of mean features extracted from the raw fNIRS signals in the t-SNE embedding space. In contrast, Figure \ref{fig6} (b) illustrates a clustering phenomenon in the distribution of various categories following the OCMCRD, although not completely separable. Specifically, fNIRS data from different subjects under the same emotional stimulus exhibit more similar feature distributions. Notably, the blue data points are farther from the orange data points in the embedding space compared to the green (or pink) data points. This validates that the IS-ICR contrastive learning strategy pulls the less similar features further apart. Overall, the proposed method effectively mitigates inter-subject differences without losing emotion separability, thereby facilitating cross-subject emotion recognition. This is beneficial for deploying the model onto wearable physiological monitoring devices to predict emotion categories for unseen subjects.

\section{Conclusion}

In this study, we propose a novel method (OMCRD) for lightweight the fNIRS emotion recognition model. The framework allows multiple students to jointly optimize the network and teach cross-subject knowledge at region and channel levels interactively. Furthermore, the IS-ICR loss boosts the similarity between identical emotional representations from various individuals, notably improving the model's capability for cross-subject adaptive recognition. Our framework shows effective emotion recognition, robustness across subjects, and deployability in extensive experiments. The proposed OMCRD is a general framework for multi-channel fNIRS sequences that can be extended to large-scale models in fields like mental health diagnosis and cognitive assessment.

\begin{acks}
This work is partially supported by the following grants:
National Natural Science Foundation of China (61972163,
U1801262), Natural Science Foundation of Guangdong Province (2022A1515011555, 2023A1515012568), Guangdong Provincial Key Laboratory of Human Digital Twin (2022B1212010004).
\end{acks}

\bibliographystyle{ACM-Reference-Format}
\balance
\bibliography{ref}










\end{document}